%% file: macro_IAUS370.tex
\documentclass{iau} 
\usepackage{graphicx}
\usepackage{natbib}
\usepackage{amsmath}
\usepackage{amssymb}
\usepackage{arydshln}
\usepackage{txfonts}
\newcommand{\sgs}{SMCSGS-FS\,69}

\input{bibdefinitions.tex}

\pubyear{2022}
\volume{370}
\jname{Winds of Stars and Exoplanets}
\editors{A.~A. Vidotto, L.~Fossati \& J.~Vink, eds.}

\title[Winds of OB stars: impact of metallicity, rotation and binary interaction] 
{Winds of OB stars: impact of metallicity, rotation and binary interaction}

\author[Varsha Ramachandran
 et al]   
{Varsha Ramachandran
$^1$}

\affiliation{$^1$Zentrum f{\"u}r Astronomie der Universit{\"a}t Heidelberg,
Astronomisches Rechen-Institut, M{\"o}nchhofstr. 12-14, 69120 Heidelberg \\ [\affilskip]}

\begin{document}

\maketitle

\begin{abstract}
Winds of massive stars are an important ingredient in determining their evolution, final remnant mass, and feedback to the surrounding interstellar medium. We compare empirical results for OB star winds at low metallicity  with theoretical predictions. Observations suggest very weak winds at SMC metallicity, but there are exceptions. We identified promising candidates for rotationally enhanced mass-loss rates with two component wind and partially stripped stars hiding among OB stars with slow but dense wind in the SMC. A preliminary analysis of these systems, derived parameters, and their implications are discussed. Finally, we briefly discuss the interaction of OB winds near black holes in X-ray binaries. 
\end{abstract}

\firstsection 
              
\section{Introduction}
 
Massive stars are hot, luminous objects that lose mass due to stellar winds driven by the scattering of UV photons by metal lines \citep[e.g.][]{CAK1975, Graefener2005}.  Stellar winds have a major role in the evolution of massive stars throughout the course of their lifetimes, as well as in determining their ultimate fate and the masses of compact remnants.
Detection of gravitational waves from coalescing black holes further highlighted the need for a better understanding of massive star winds, especially at low metallicity.
 
The winds of hot stars are characterized by two main parameters: the terminal velocity $\varv_\infty$ and the rate of mass-loss $\dot{M}$. We see direct evidence of mass-loss in the UV spectral line profiles of highly ionized species such as \ion{C}{iv}, \ion{Si}{iv}, and \ion{N}{v}. Radiatively-driven winds are by nature
metallicity dependent. Moreover, fast rotation and binary interaction can have a strong impact on the stellar wind and the evolution of the massive star. In this study, we highlight empirical studies of OB stars at low metallicity, focusing on their wind properties. Furthermore, we discuss the influence of rapid rotation and binary interaction on the wind properties.

\section{Weak winds of massive stars at low metallicity}
\label{sect:weakwind}

The spectroscopic analysis using stellar atmosphere models enables us to quantify wind mass-loss rates. UV spectroscopy along with optical observations are particularly suitable for this. Previous theoretical and empirical studies suggest that mass-loss rates depend on Z$^\alpha$, with $\alpha$ between 0.7 and 0.8 \citep{Vink2001,Mokiem2007}. 
 However, there is growing evidence for lower mass-loss rates for OB stars \citep{Bouret2003,Martins2004} compared to widely used theoretical recipes \citep{Vink2000,Vink2001}. Lower mass-loss rates are observed  for Galactic massive stars with luminosities less than $\log(L/L_\odot)\lesssim5.2 $ (late O and early B-type dwarfs), also known as the ``weak-wind problem'' \citep{Martins2004,Puls2008}. However, empirical determination of mass-loss rates of OB stars at SMC metallicity was sparse before. The major reasons include the absence of optical emission lines in the optical spectra due to low metallicity and unavailability of UV spectra.

\begin{figure}[h]
\begin{center}
 \includegraphics[width=0.7\textwidth]{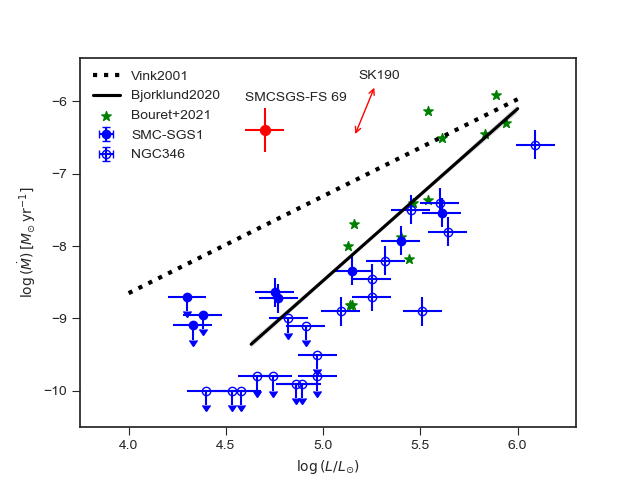} 
 \caption{Mass-loss rate as a function of stellar
luminosity for OB stars in SMC \citep{Ramachandran2019, Rickard2022}.
 For comparison, the theoretical predictions from \cite{Vink2000,Vink2001}  (dashed black line),  and \cite{Bjorklund2021} (solid black line) corresponding to SMC metallicities are marked. Mass-loss rates of the stripped star and rapid rotator are marked in red and  giants/supergiants in the SMC are shown in green asterisks for comparison \citep{Bouret2021}. }
 \label{fig:wlr} 
\end{center}
\end{figure}

 Figure\,\ref{fig:wlr} depicts the mass-loss rate ($\dot{M}$) versus luminosity for SMC OB stars. The samples consist of non-supergiant OB stars in the Wing of the SMC \cite{Ramachandran2019} and in the young massive cluster NGC\,346 \cite{Rickard2022}. The spectral analyses of these samples were performed using PoWR\footnote{http://www.astro.physik.uni-potsdam.de/$\sim$wrh/PoWR} model atmospheres. A linear regression to this $\log \dot{M} - \log L$ relation, which accounts for the individual error 
bars, shows a steeper relation and more than an order of magnitude offset across all luminosity ranges compared to the 
theoretical predictions \citep{Vink2001} at SMC metallicity.  The results are consistent with earlier spectroscopic investigations of massive stars in the SMC \citep{Bouret2003,Martins2004} and other low-metallicity dwarf galaxies like IC 1613 and WLM \citep{Bouret2015,Lucy2012}. Although the empirical results are close to  \cite{Bjorklund2021}, they still overestimate the mass-loss rates by a factor of two or more. It also to be noted that the mass-loss rates of luminous supergiants in the SMC \citep{Bouret2021} are found to be in agreement with \citet{Vink2001} predictions unlike OB dwarfs and giants.

The weak winds of OB stars at low metallicity should only have a small influence on their evolution. The fact that theoretical evolutionary tracks are typically constructed using conventional mass-loss recipes, which are significantly overestimated compared to observations, may modify our understanding of massive star evolution. This also affects the gravitational wave population synthesis and stellar feedback in galaxy evolution simulations. We have secured HST UV spectroscopy of low metallicity O stars in the Magellanic Bridge which will be further studied in detail to understand the stellar winds of sub-SMC metallicity stars \citep{Ramachandran2021}.

\section{Stellar wind in rapidly rotating stars}

Rotation modifies the shape of the star, deviating from spherical symmetry and thereby influencing stellar parameters such as effective gravity, effective temperature, and radiative flux. The winds in fast rotating massive stars do not remain isotropic but become increasingly anisotropic as the rotation approaches the critical rotation \citep{Maeder2000ARA}. Very rapid rotators are expected to have cooler photospheres closer to their equatorial regions (due to gravity darkening), and that the local mass loss rate attains a minimum at the equator and the strongest winds occur at the poles \citep{Owocki1996, Maeder2000ARA,Muller2014}. However if the temperature at the equator reaches below the bistability jump due to gravity darkening,  mass-loss rate might be enhanced over equatorial regions \citep{Lamers1991}. In summary, latitude dependent temperature and gravity by fast rotation are expected to result in asymmetric winds and enhanced mass-loss rates.

Few such systems have been studied in the Galaxy \citep{Bjorkman1994, Massa1995, Prinja1997} and in the LMC \citep{Shepard2020}. However no such fast rotating OB stars with rotationally enhanced wind were reported in the SMC. One exception is the luminous O supergiant AV 83, showing  extreme wind mass-loss rate despite slow rotation velocity and no asymmetric wind line profiles \citep{Hillier2003}. 
Investigating rotationally enhanced mass-loss rate at low metallicity is of foremost importance since most of the OB stars have very weak winds whereas fast rotation is common. 

We found one such good candidate, namely SK\,190, an O giant/ supergiant star in the Wing of the SMC. Based on the optical spectral analysis using PoWR, we derived the stellar parameters of the star to be $T_\ast$ = 
30\,kK, log\,$g_\ast$ = 3.2, log\,$L/L_{\odot}$\,=\,5.3 and $\varv$\,sin\,$i$=300 km $s^{-1} $ $\sim 0.8 \varv_{\mathrm{crit}}$. However, the UV lines show evidence for latitude dependent parameters: a hot pole and a cool equator with $\approx6$\,kK difference in temperature, higher surface gravities at the pole by $\approx0.4$\,dex and two wind components.  Saturate UV P-Cygni profiles \ion{C}{iv} and \ion{Si}{iv} show a narrow absorption component and  a broad emission part, suggesting that the terminal wind velocity increases from equator to poles (Figure\,\ref{fig:wlr}). In the spectroscopic analysis we had to assume a slow wind of 400 km\,$s^{-1}$ to account for the narrow absorption and a fast wind of 1600 km $s^{-1}$ to fit the broad emission. In addition high ionization lines such \ion{N}{v} and \ion{O}{v} present in the spectra can be only reproduced using hotter temperature models than those used to fit optical spectra. Based on models assuming cooler temperature, lower surface gravity and slow wind velocity  we estimated a mass-loss rate of $\log\dot{M} = -6.4 M _{\odot}\,\mathrm{yr^{-1}} $ at the equator whereas hotter, higher surface gravity  models corresponding to the pole with faster wind velocities suggested a much higher mass-loss rate  $\log\dot{M} = -5.8 M_{\odot}\,\mathrm{yr^{-1}} $. The mass-loss rate constraints for this star is compared with that of OB stars and supergiants in the SMC in Figure\,\ref{fig:wlr}. Only a few luminous supergiants in the SMC show such high mass-loss rates. Moreover, stars of similar luminosity as that of Sk\,190 show many orders of magnitude lower mass-loss rates.

\begin{figure}[h]
\begin{center}
 \includegraphics[width=0.47\textwidth]{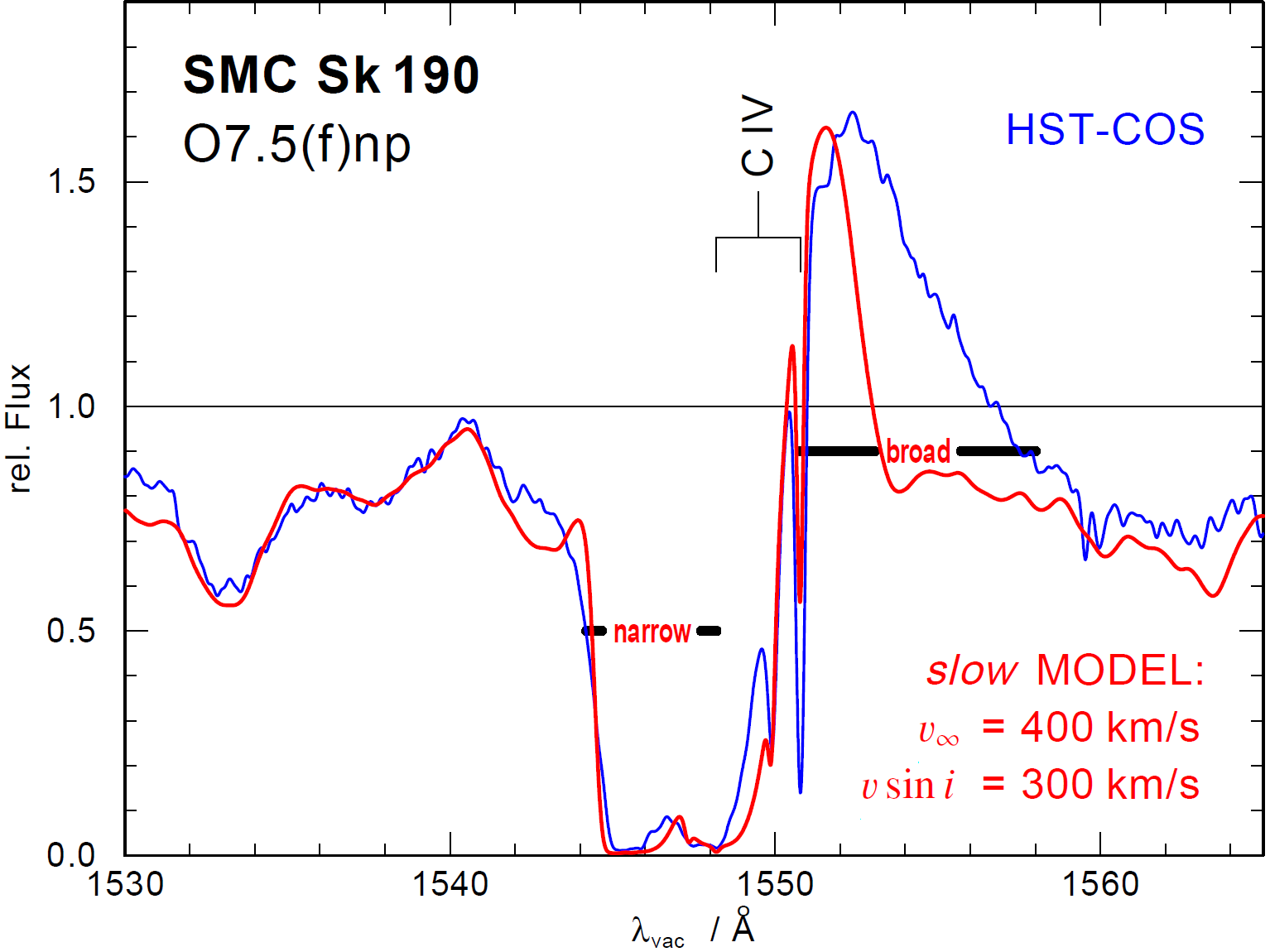} 
  \includegraphics[width=0.47\textwidth]{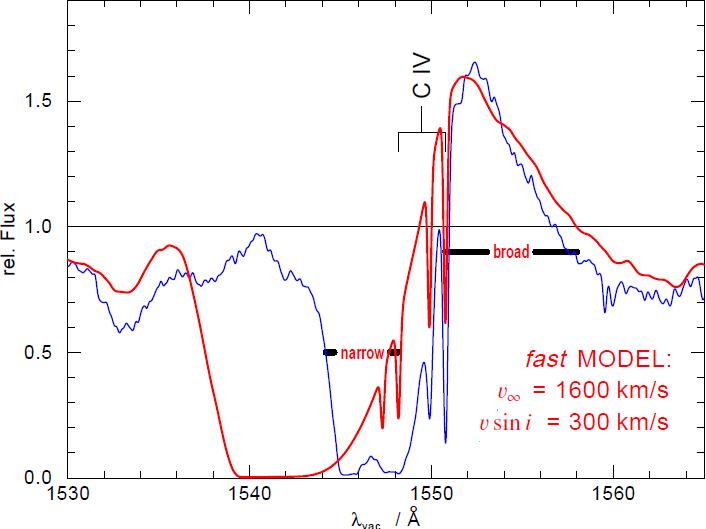} 
 \caption{The \ion{C}{iv} UV wind profile of Sk\,190 (blue) compared with models calculated with hot, fast and dense wind corresponding to the poles ($T_\ast$ = 
36\,kK, $\varv_\infty$=1600 km $s^{-1} $, $\log\dot{M} = -5.8 M_{\odot}\,\mathrm{yr^{-1}} $ ) and cool slow wind at the equator ($T_\ast$ = 
30\,kK, $\varv_\infty$=400 km $s^{-1} $, $\log\dot{M} = -6.4 M_{\odot}\,\mathrm{yr^{-1}} $ ).} \label{fig:civ} 
\end{center}
\end{figure}

\section{Stellar wind in stripped star}
Interactions in OB binaries can often result in the stripping of the primary's envelope \citep{Paczynski1967}, generating hot and compact helium core stars with only a thin layer of hydrogen on top \citep[e.g.,][]{Yoon2010,Claeys2011}. Depending on their initial masses, the stripped envelop primaries would have spectral characteristics ranging from hot subdwarf to Wolf-Rayet (WR) stars \citep{Paczynski1967,Vanbeveren1991,Eldridge2008,gotberg_2017_ionizing}. On the other hand, the secondaries would become rapidly rotating stars \citep{deMink2013}, showing disk emission features like Be stars \citep{Shao2014}.

\begin{figure}[h]
\begin{center}
\includegraphics[width=\textwidth,trim={0 0cm 0 6.6cm},clip]{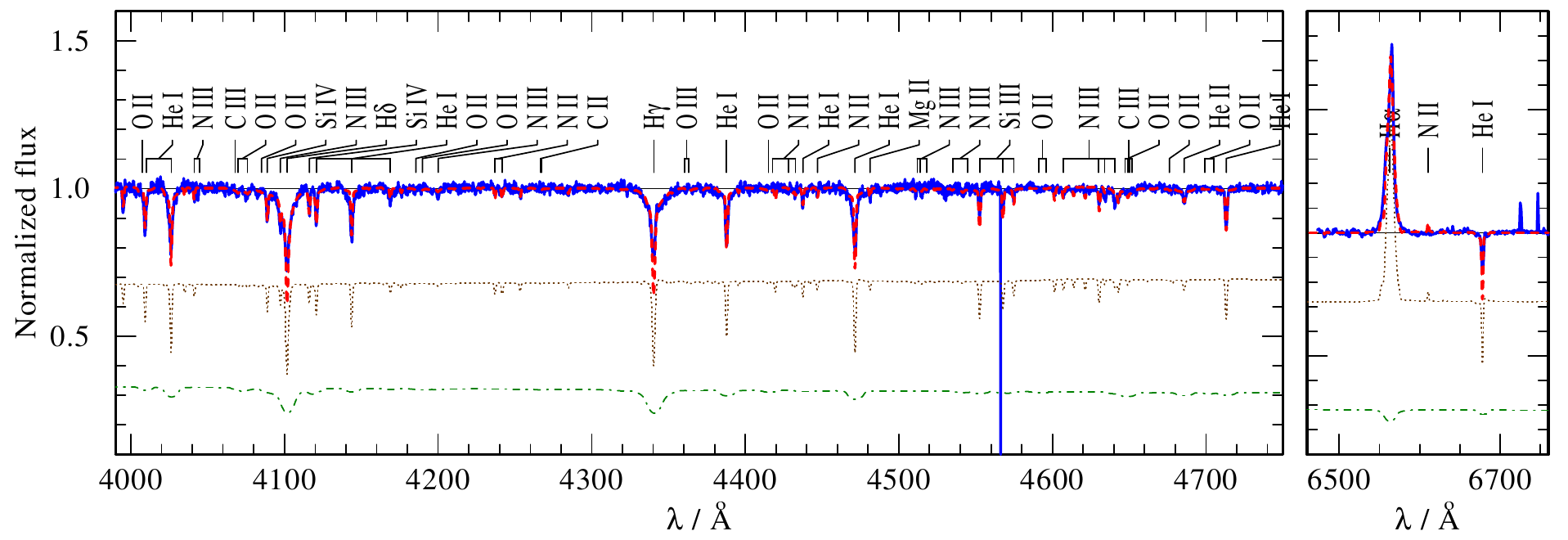}
\caption{Comparison of observed spectra (blue) with composite PoWR models (red) for \sgs. The final model is a combination of stripped primary (brown) and fast rotating secondary (green).}
\label{fig:opt}
\end{center}
\end{figure}
 
Interestingly, stripped stars with masses in between classical WR stars and low-mass subdwarfs are rarely observed. The only known candidate for an intermediate-mass He stars is the so-called qWR star HD 45166 \citep{Groh2008} with other known stripped stars being in the lower mass range of $<1.5M_{\odot}$ and the appearance of subdwarfs \citep{Wang2021,Schootemeijer2018}. Some of the recently proposed X-ray quiet BH + Be binaries such as LB1 and HR6819 \citep{Liu2019,Rivinius2020} have been later disputed \citep[e.g.,][]{Bodensteiner2020,Shenar2020} or confirmed \citep{Frost2022} to be a stripped star + Be star binary. However, both are recently detached systems in which the stripped star appears as a partially stripped cool supergiant. Recent state-of-the-art evolutionary models predict that metallicity has a substantial effect on the course and outcome of mass transfer evolution of massive binaries, leading to a large fraction of such partially stripped donors at low metallicity \citep{Klencki2022}. These systems are predicted to be hidden among OB binaries as main sequence or supergiant stars. 

We identified a partially stripped star in a binary for the first time at low metallicity, \sgs\, (Ramachandran et al. in prep). A careful analysis of the optical spectra of \sgs\ revealed the presence of a narrow-lined and broad-lined star (see Figure\,\ref{fig:opt}). The narrow line star has a slowly rotating B supergiant-like spectral features but with very low luminosity (log\,$L/L_{\odot}$\,=\,4.7). The star is likely to be core-He burning, showing very high nitrogen enrichment, enhanced He and CO depletion. The spectroscopic mass of the star is only $\lesssim$3\Msun\ making it one of the first reports of a partially stripped intermediate-mass star. The mass gainer secondary is found to be a rapidly rotating main sequence B star. As shown in Figure\,\ref{fig:opt} (left), the $H\alpha$ is in strong emission. By reproducing this spectral feature assuming it is formed in the wind of the stripped star, we suggest an upper-limit for the mass-loss rate to be $\log\dot{M} = -6.2 M _{\odot}\,\mathrm{yr^{-1}} $. However, the $H\alpha$ emission could be a composite of wind from the stripped star and disk emission from the fast rotating secondary, suggesting $\log\dot{M} $ could be lower ($\lesssim-6.6 M _{\odot}\,\mathrm{yr^{-1}} $). Nevertheless, the mass-loss rate of the stripped star is significantly higher than that of OB stars of the same luminosity (Figure\,\ref{fig:wlr}). Similar systems with high mass-loss rates may be lurking among ordinary OB stars, where they will be significant contributors to stellar wind feedback.

\section{Stellar wind of OB star in black hole X-ray binary}

To understand the complex behaviour of High Mass X-ray binaries (HMXBs) with black hole (BH) companions, detailed knowledge of the massive star donors is essential. However, only a few such systems
are known so far. To remedy this situation, we performed a multi-wavelength phase-resolved analysis of the extragalactic HMXB M33 X-7. This eclipsing BH HMXB is reported to contain a very massive O supergiant donor and a massive black hole in a short orbit \citep{Orosz2007,Pietsch2006}.  However, previous spectroscopic analyses were limited to plane-parallel models which are optimized for hot stars with no significant wind.

\begin{figure}
\begin{center}
\includegraphics[width=0.85\textwidth,trim={0cm 0cm 0 0cm},clip]{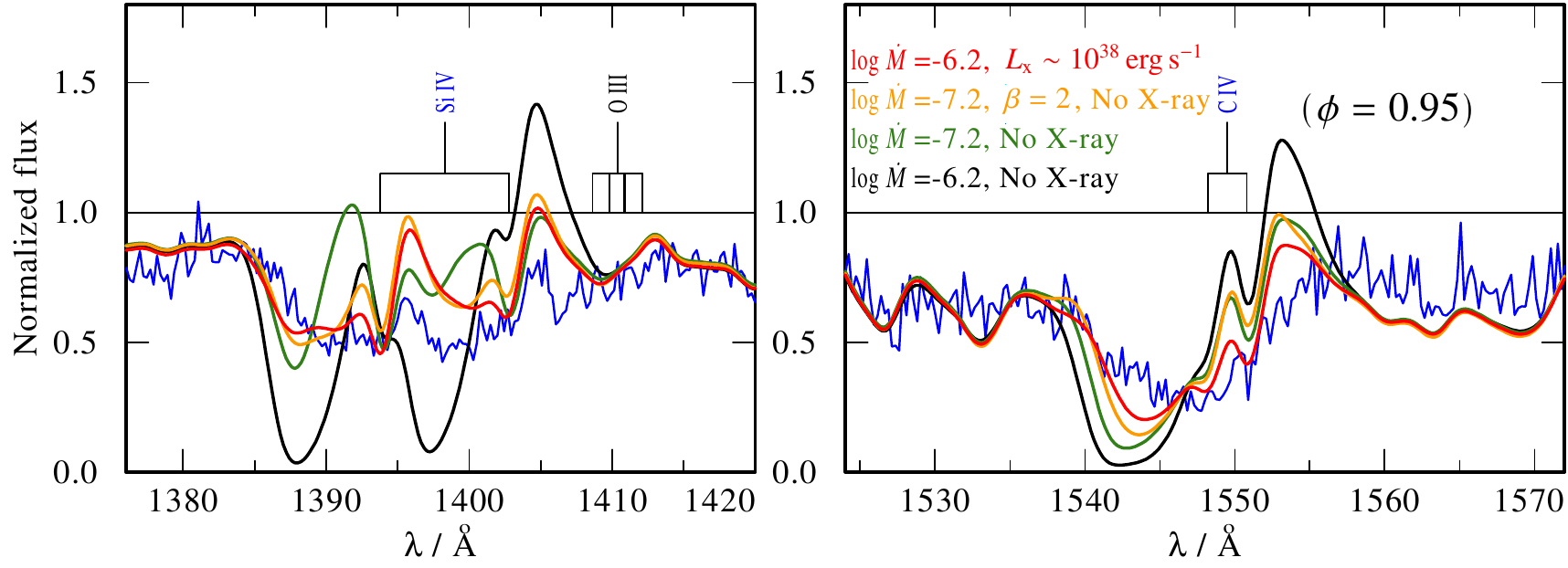} 
\includegraphics[width=0.85\textwidth]{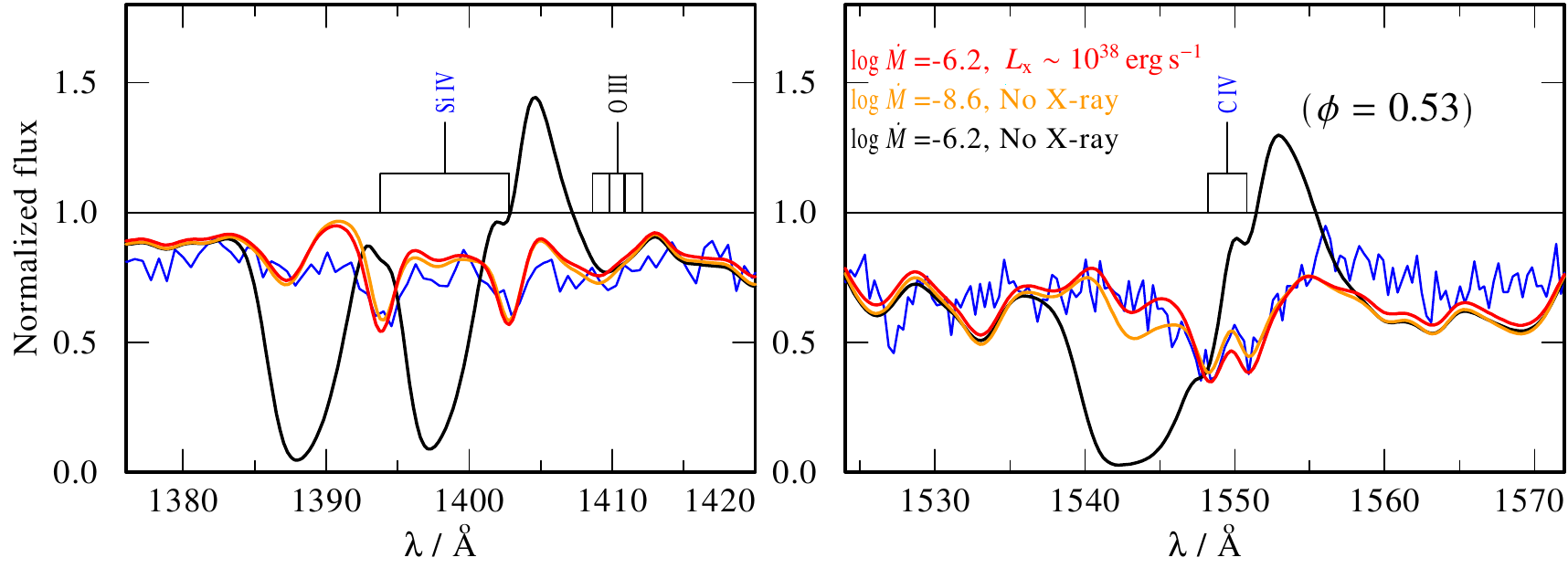} 
\caption{Observed \ion{Si}{iv} (left and \ion{C}{iv} (right) wind lines (blue solid) of M33 X-7 taken at eclipse (upper) and inferior conjunction (lower) compared to the models  with different wind parameters and X-rays.}
\label{fig:m33x7}
\end{center}
\end{figure} 

Using phase-resolved simultaneous \hst- and \xmm-observations, we trace the interaction of the stellar wind with the BH. 
The UV resonance lines show the Hatchett-McCray effect with a large reduction in absorption strength when the BH is in the foreground due to the strong X-ray ionization (see Figure\,\ref{fig:m33x7}).
Our comprehensive spectroscopic investigation of the donor star (X-ray+UV+optical) yields new stellar and wind parameters for the system that differ significantly from previous estimates \citep[see][for more details]{Ramachandran2022}. In particular, the masses of the components are considerably reduced to $\approx 38M_{\odot}$ for the O-star donor and $\approx 11.4 M_{\odot}$ for the black hole. The O giant is overfilling its Roche lobe and shows surface He enrichment.

The derived mass-loss rate of the donor is in good agreement with the \cite{Vink2001} prediction assuming a high depth-dependent microclumping. By incorporating observed X-ray luminosities in models corresponding to  different orbital phases, we were able to reproduce the spectral variations at three phases  with the same stellar and wind parameters. 
We investigated the wind driving contributions from different ions and the changes in the ionization structure due to X-ray illumination. Towards the black hole, the wind is strongly quenched due to strong X-ray illumination. For this system, the standard wind-fed accretion scenario alone cannot explain the observed X-ray luminosity, pointing towards an additional mass overflow, in line with our acceleration calculations. The classical distinction between wind-fed and Roche-lobe overflow systems becomes meaningless for our system. 
Our investigations on wind driving and the impact of X-rays in M33 X-7 can be also applied to other high luminosity HMXB systems in general.


\def\apj{{ApJ}}    
\def\nat{{Nature}}    
\def\jgr{{JGR}}    
\def\apjl{{ApJ Letters}}    
\def\aap{{A\&A}}   
\def\mnras{{MNRAS}}
\def\aj{{AJ}}
\let\mnrasl=\mnras


\bibliographystyle{aa}
\bibliography{ref}

\end{document}

%% file: bibdefinitions.tex
\DeclareRobustCommand{\ion}[2]{%
	\relax\ifmmode
	\ifx\testbx\f@series
	{\mathbf{#1\,\mathsc{#2}}}\else
	{\mathrm{#1\,\mathsc{#2}}}\fi
	\else\textup{#1\,{\mdseries\textsc{#2}}}%
	\fi}

\newcommand{\xmm}{{\em XMM-Newton}}

\newcommand{\hst}{{\em HST}}

\newcommand{\Msun}{\mbox{$M_\odot$}\ }

\def\aj{AJ}%
\def\apj{ApJ}%
\def\apjl{ApJ}%
\def\aap{A\&A}%
\def\mnras{MNRAS}%
\def\nat{Nature}%
\def\jgr{J.~Geophys.~Res.}%